\documentclass{paper}
\usepackage{epsfig, graphicx}
\usepackage[english]{babel}

\begin{document}
\title{Solid –- Liquid Phase Transition in a Gibbs Monolayer of Melissic Acid at the n-Hexane –- Water Interface}
\author{\small Aleksey M. Tikhonov\/\thanks{tikhonov@kapitza.ras.ru}}
\maketitle
\leftline{\it Kapitza Institute for Physical Problems, Russian Academy of Sciences,}
\leftline{\it ul. Kosygina 2, Moscow, 119334, Russia}

\rightline{\today}

\abstract{A sharp phase transition from a crystalline state with the area per molecule $A = (17 \pm 1)$~\AA$^2$ to a liquid state with $A = (23 \pm 1)$~\AA$^2$ at the n-hexane –- water interface in a Gibbs monolayer of melissic acid has been revealed in data of X-ray reflectometry with the use of synchrotron radiation.}

\vspace{0.25in}

\large

The solid –- liquid phase transition on the surface of
a high-molecular alkane is observed above the bulk
melting temperature \cite{1,2}. The possibility of such a
phase transition at an alkane–water interface has been
discussed for a long time \cite{3,4,5,6,7}. The observation of a
solid –- liquid phase transition at the n-hexane –- water
interface in the Gibbs monolayer of a surfactant is
reported in this work.

Melissic acid (C$_{30}$-acid) is not solved in water and is
weakly solved in n-hexane (C$_6$H$_{14}$, density $\approx 0.65$\,g/cm$^3$
at 298\,K, boiling temperature $T_b \approx 342$\,K), which is
hardly solved in water under normal conditions. At
quite low temperatures, C$_{30}$H$_{60}$O$_2$ molecules of acid are
adsorbed from a solution in a hydrocarbon solvent on
the n-hexane -– water interface in the form of a solid
monolayer (Gibbs monolayer) with the thermodynamic
parameters $(p, T, c)$, significantly reducing its
energy \cite{8}. According to our new data, with an
increase in the temperature $T$ (at the pressure $p = 1$\,atm), a phase
transition occurs in the monolayer at a
temperature $T_c$ that is determined by the concentration
$c$ of the C$_{30}$-acid in the volume of the solvent serving
as a reservoir for the surfactant molecules.

A sample of a macroscopically flat n-hexane –-
water interface oriented by the gravitational force was
studied in a stainless-steel cell (see Fig. 1). The dimensions
of the interface were $75 \times 150$\,mm \cite{9}. The surface
tension of the interface $\gamma(T)$ was measured by the
Wilhelmy plate method with the cell placed in a homemade
single-stage thermostat \cite{10}. To this end, through holes
with a diameter of $\sim 1$\,mm were made in its upper cap
and in the cap of the hatch of the cell. The reflectometry
of the n-hexane -– water interface was performed in
the hermetically closed cell and its temperature $T$ was
controlled in a homemade two-stage thermostat. For
the entrance and exit of X rays, as well as for the convenient
visual observation of the interface, the windows
of the cell were fabricated from transparent polyester
(Mylar).

\begin{figure}
\hspace{0.5in}
\epsfig{file=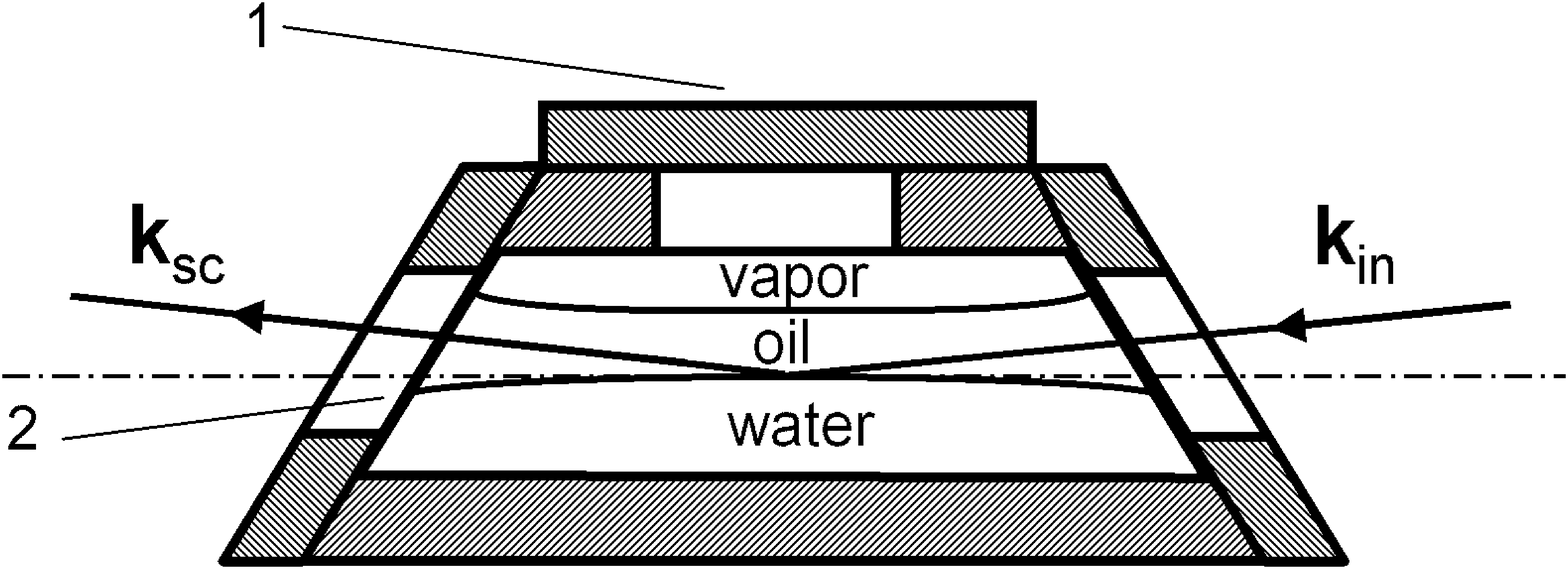, width=0.75\textwidth}

Figure 1. Airtight sample cell for study of the planar liquid –- liquid
interface: (1) the removable cap of the hatch and (2) the
transparent polyester (Mylar) window.
\end{figure}

A solution of sulfuric acid (ðÍ = 2) in deionized
water (Barnstead, NanoPureUV) with a volume of
about 100\,mL was used as the lower bulk
phase. The solution of melissic acid in n-hexane with
a volume of about 100\,mL and the volume
concentration $c \approx 0.2$\,mmol/kg $(\approx 2 \cdot 10^{-5})$ was used
as the upper bulk phase. C$_{30}$-acid and n-hexane were
purchased from the Sigma-Aldrich Corporation. The
alkane was preliminarily purified by multiple filtration
in a chromatography column through a thick ($\sim 30$\,cm)
layer of a fine-grained aluminum oxide powder with a
particle diameter of $\sim 0.1$\,mm. C$_{30}$-acid was doubly
purified by recrystallization at room temperature from
a supersaturated solution in n-hexane, which was prepared
by the solution of the acid in n-hexane at a temperature
of $T \approx 333$\,K \cite{11}. Before the measurements
of the reflection coefficient $R$ from the interface, the
sample was "annealed": the liquids in the cell were
heated to $T \approx T_b$ and were then cooled below $T_c$. Thus,
the formation of gas bubbles at the n-hexane -– water
interface at a change in $T$ was prevented in subsequent
experiments.

\begin{figure}
\hspace{0.5in}
\epsfig{file=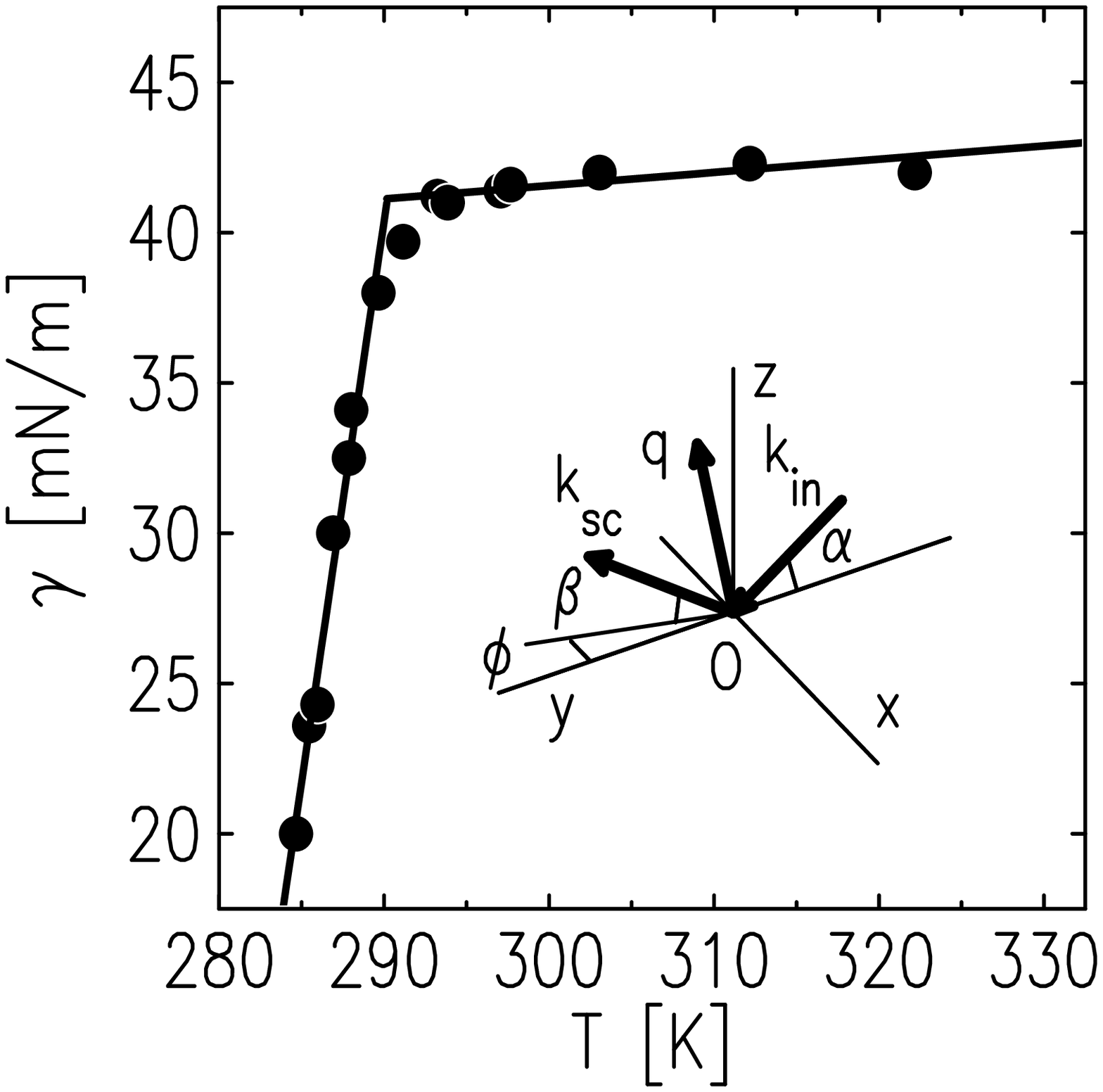, width=0.7\textwidth}

Figure 2. Temperature dependence of the surface tension of
the n-hexane -– water interface at the concentration $c \approx
0.2$\,mmol/kg of C$_{30}$-acid in n-hexane. The bending point
corresponds to $T_c \approx 291$\,K. The inset shows the kinematics
of scattering at the n-hexane -– water interface. The $(xy)$ plane
coincides with the interface; the $Ox$ axis is perpendicular to
the beam direction; the $Oz$ axis is normal to the surface and
is opposite to the gravitational force; {\bf k}$_{\rm in}$ and {\bf k}$_{\rm sc}$ are the
wave vectors of the incident and scattered beams in the
direction of the observation point, respectively; {\bf q = k$_{\rm in}$ {\rm -} k$_{\rm sc}$}
is the scattering vector; and $\alpha$ and $\beta$ are the grazing and
scattering angles in the plane normal to the surface.
\end{figure}

A Wilhelmy plate made of a chromatographic paper
(Wattman) with a length of $L \approx 10$\,mm and a width of
$\approx 5$\,mm was used in the measurements of $\gamma(T)$. It was
fastened to a thin (diameter $\sim 0.25$\,mm) platinum wire
passing through holes in the caps of the thermostat
and hatch of the cell (see Fig. 1). The maximum
change in the weight of the plate $\Delta F$ was fixed by an
electric balance (NIMA PS-2) at its slow pulling from
the lower phase. Figure 2 shows the dependence
$\gamma(T)\approx \Delta F/2L$, which exhibits a feature (kink) at the
phase transition temperature $T \approx 291$\,K. A change in
the slope of $\gamma(T)$ is related to a change in the surface
enthalpy $\Delta H = - T_c\Delta(\partial \gamma/\partial T)_{p,c}$  $=1.1\pm 0.1$\,J/m$^2$.

Methods based on the scattering of synchrotron
X-ray radiation currently provide the main information
on the microscopic structure of (nonpolar
organic solvent -– water) interfaces, which cannot be
obtained from the measurements of such characteristics
as the surface tension, capacity of the interface,
and surface potential \cite{10,12}. Unfortunately, relatively
strong scattering in the bulk of the hydrocarbon
solvent at $\lambda \sim 1$\,\AA{} prevents the application of the grazing
diffraction method to study the in-plane crystal
order at the liquid -– liquid interface. The transverse
structure of the interface was studied by X-ray reflectometry
with the use of synchrotron radiation at the
X19C station of the NSLS synchrotron, which was
equipped with a universal spectrometer for studying
the surface of the liquid \cite{13}. A bending magnet with
a critical energy of  $\sim 6$\,keV was a source of radiation for
the X19C station. In experiments, a focused monochromatic
beam with an intensity of $\approx 10^{11}$ photons/s
and a photon energy of $E=15$\,keV ($\lambda=0.825 \pm 0.002$ \,\AA) was used.

\begin{figure}
\hspace{0.5in}
\epsfig{file=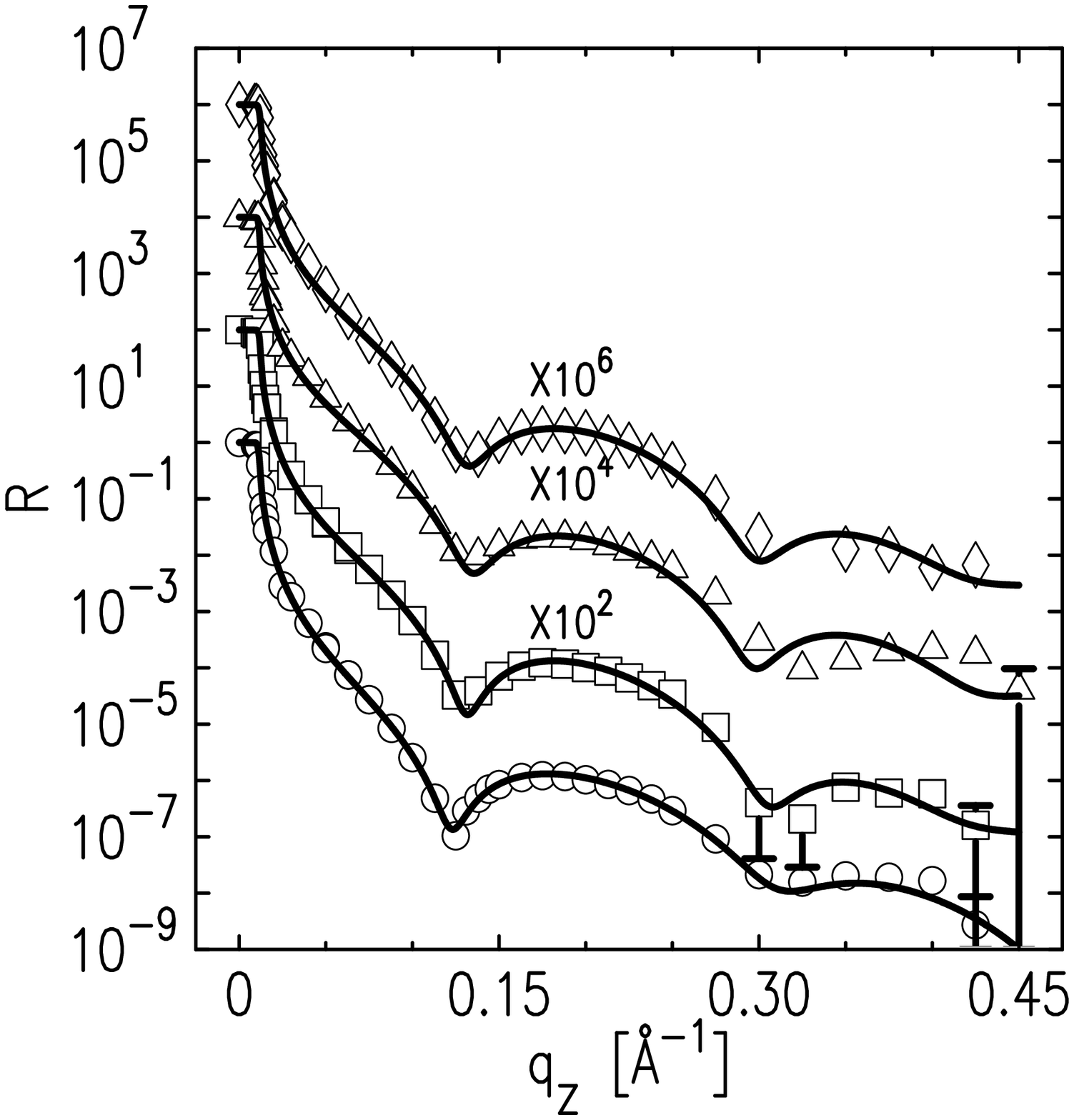, width=0.7\textwidth}

Figure 3. Reflection coefficient $R$ versus $q_z$ for the n-hexane -–
water interface at (diamonds) 293.4, (triangles) 290.1,
(squares) 289.2, and (circles) 288.3\,K. The solid lines correspond
to the two-layer model of the absorbed monolayer.

\end{figure}

Since the n-hexane –- water interface is oriented by
the gravitational force, the scattering kinematics is
conveniently described in the right-handed Cartesian
coordinate system whose origin $O$ is at the center of
the illuminated region, the $(xy)$ plane coincides with
the interface between the monolayer and water, the
$Ox$ axis is perpendicular to the beam direction, and the
$Oz$ axis is normal to the surface and is opposite to the
gravitational force (see the inset in Fig. 2). Let {\bf k}$_{\rm in}$ 
and {\bf k}$_{\rm sc}$ be the wave vectors of the incident and scattered
beams in the direction of the observation point,
respectively. In the case of mirror reflection, $\alpha = \beta$ and
$\phi = 0$, where $\alpha$ is the grazing angle in the $(yz)$ plane, $\beta$
is the angle between the scattering direction and the
interface in the vertical plane, and $\phi$ is the angle
between the incident and scattered beams in the $(xy)$
plane. The scattering vector {\bf q = k$_{\rm in}$ {\rm -} k$_{\rm sc}$} at mirror
reflection has only one nonzero component $q_z=(4\pi/\lambda)\sin(\alpha)$.

\begin{figure}
\hspace{0.5in}
\epsfig{file=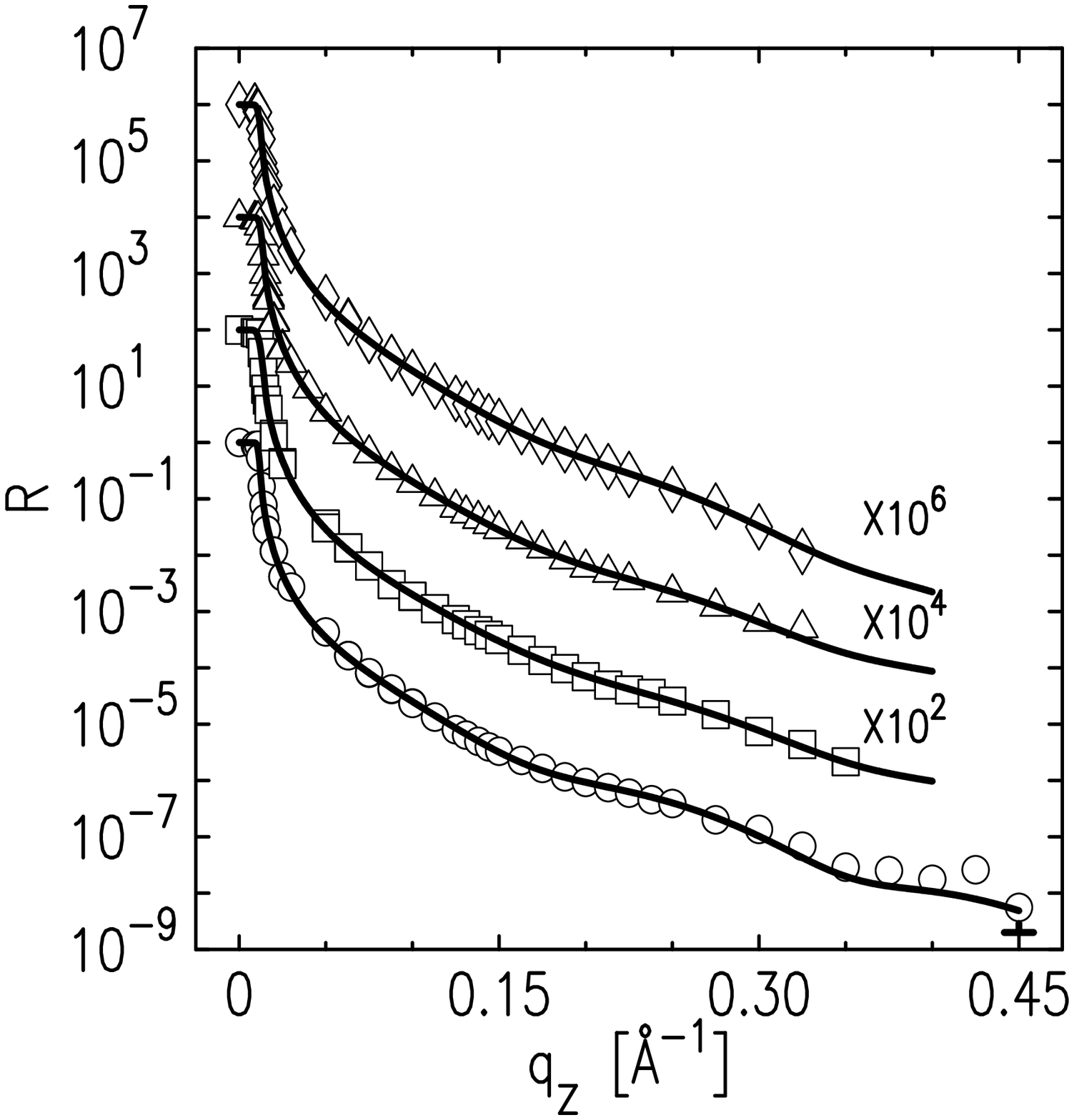, width=0.7\textwidth}

Figure 4. Reflection coefficient $R$ versus $q_z$ for the n-hexane -–
water interface at (diamonds) 334.2, (triangles) 317.9,
(squares) 308.1, and (circles) 298.2\,K. The solid lines correspond
to the two-layer model of the absorbed monolayer
described in the main text.

\end{figure}

\begin{figure}
\hspace{0.5in}
\epsfig{file=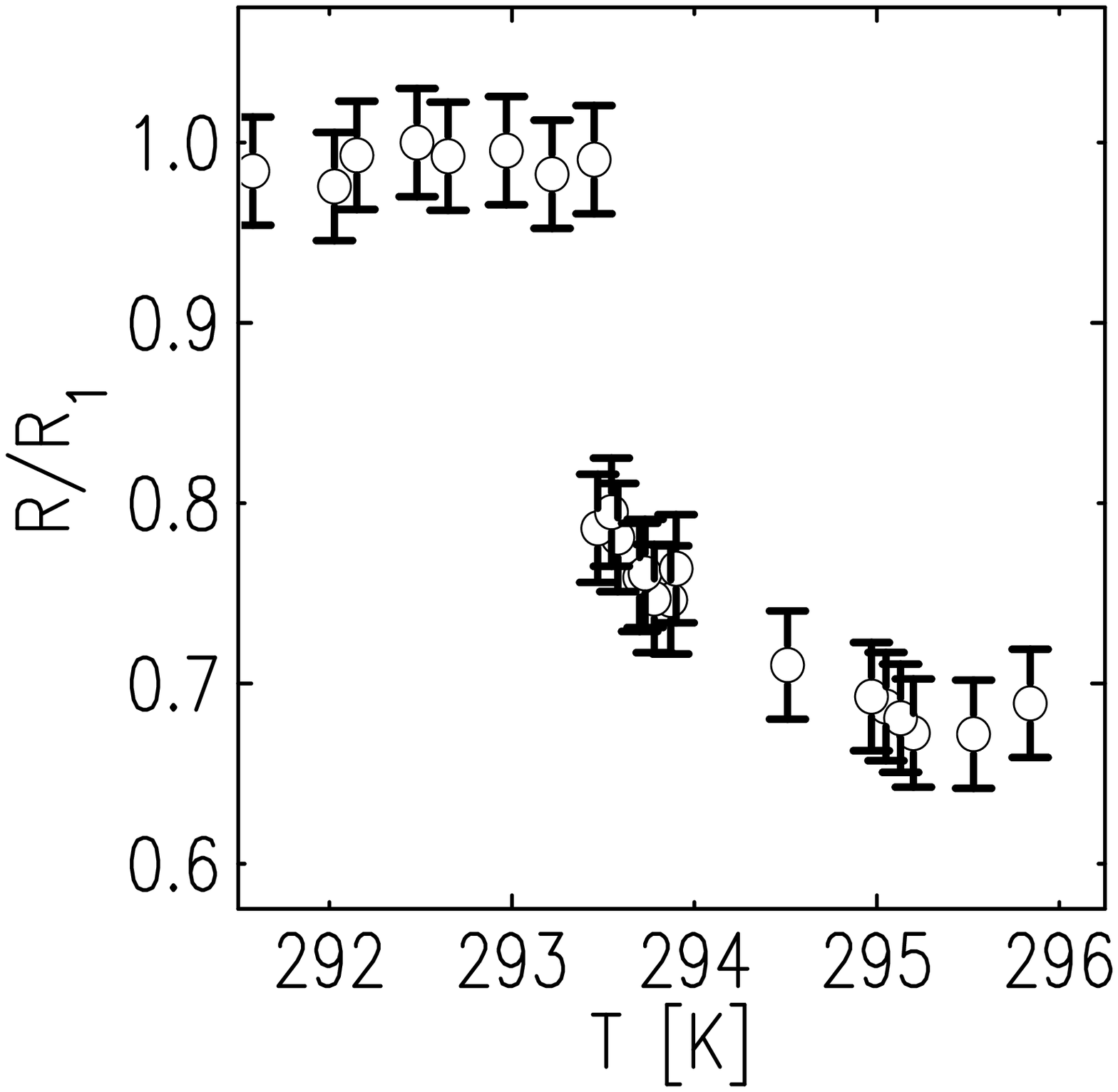, width=0.7\textwidth}

Figure 5. Temperature dependence of the normalized reflection
coefficient $R/R_1$ at $q_z=0.05$\,\AA$^{-1}$, where $R_1$ is the
reflection coefficient at $T \approx 292.2$\,Ê.
\end{figure}

The dependence of the reflection coefficient $R$ on
$q_z$ contains information on the electron density distribution
$\rho(z)$ across the n-hexane -– water interface averaged
over a macroscopic area of the illumination
region ($\sim 100$\,mm$^2$). The measurements of $R$ at low $q_z$
values are restricted by the transverse dimension and
natural divergence ($\sim 10^{-4}$\,rad) of a synchrotron radiation
beam incident on the sample. The distance
between the center of the cell and the nearest slits
limiting the vertical dimension of the incident beam
is $\sim 120$\,mm. At the smallest grazing angles of $\sim 6\cdot 10^{-4}$\,rad
($q_z\approx 0.01$\,\AA$^{-1}$), the vertical dimension of the
beam should be $\approx 15$\,$\mu$m for the illumination region to
not exceed the flat segment of the interface ($\sim 20$\,mm).
This can be achieved only by suppressing the natural
divergence of the beam to $\sim 10^{-5}$\,rad, e.g., by means of
two entrance slits with a dimension of $\sim 10$\,$\mu$m at a distance
of $\sim600$\,mm. At large grazing angles ($q_z>0.2$\,\AA$^{-1}$),
the maximum vertical dimension of the entrance slits,
0.4\,mm, is limited by the chosen vertical angular resolution
of the detector, $2\Delta\beta\approx 10^{-3}$\, (a slit with a vertical
dimension of 0.8\,mm at a distance of $\approx680$\,mm
from the center of the sample). The measurements
were performed with the resolution $\Delta\phi\approx 10^{-2}$\,rad of
the detector in the horizontal plane.

Figures 3 and 4 show the dependences $R(q_z)$ for the
n-hexane -– water interface at various temperatures
above and below the phase transition, respectively. At
$q_z < 4\pi/\lambda)\alpha_c \approx 0.01$\,\AA$^{-1}$, the incident beam undergoes
the total external reflection $R\approx 1$. The critical angle $\alpha_c$
is determined by the difference 
$\Delta\rho\approx0.11$\,{\it e$^-$/}{\AA}$^3$, 
between the volume electron densities of n-hexane
and water: $\alpha_c =\lambda\sqrt{r_e\Delta\rho/\pi}\approx 10^{-3}$\,rad, where $r_e =2.814\cdot10^{-5}$\,\AA{} is the classical radius of the electron.
The data presented in Figs. 3 and 4 clearly demonstrate
that the reflection curve changes sharply near
$T_c$. In addition to the dependences $R(q_z)$, we measured
the temperature dependence of the reflection coefficient
near $T_c$ at a fixed value of $q_z=0.05$\,\AA$^{-1}$ (see
Fig. 5) with the same spatial resolution of the detector.

\begin{figure}
\hspace{0.5in}
\epsfig{file=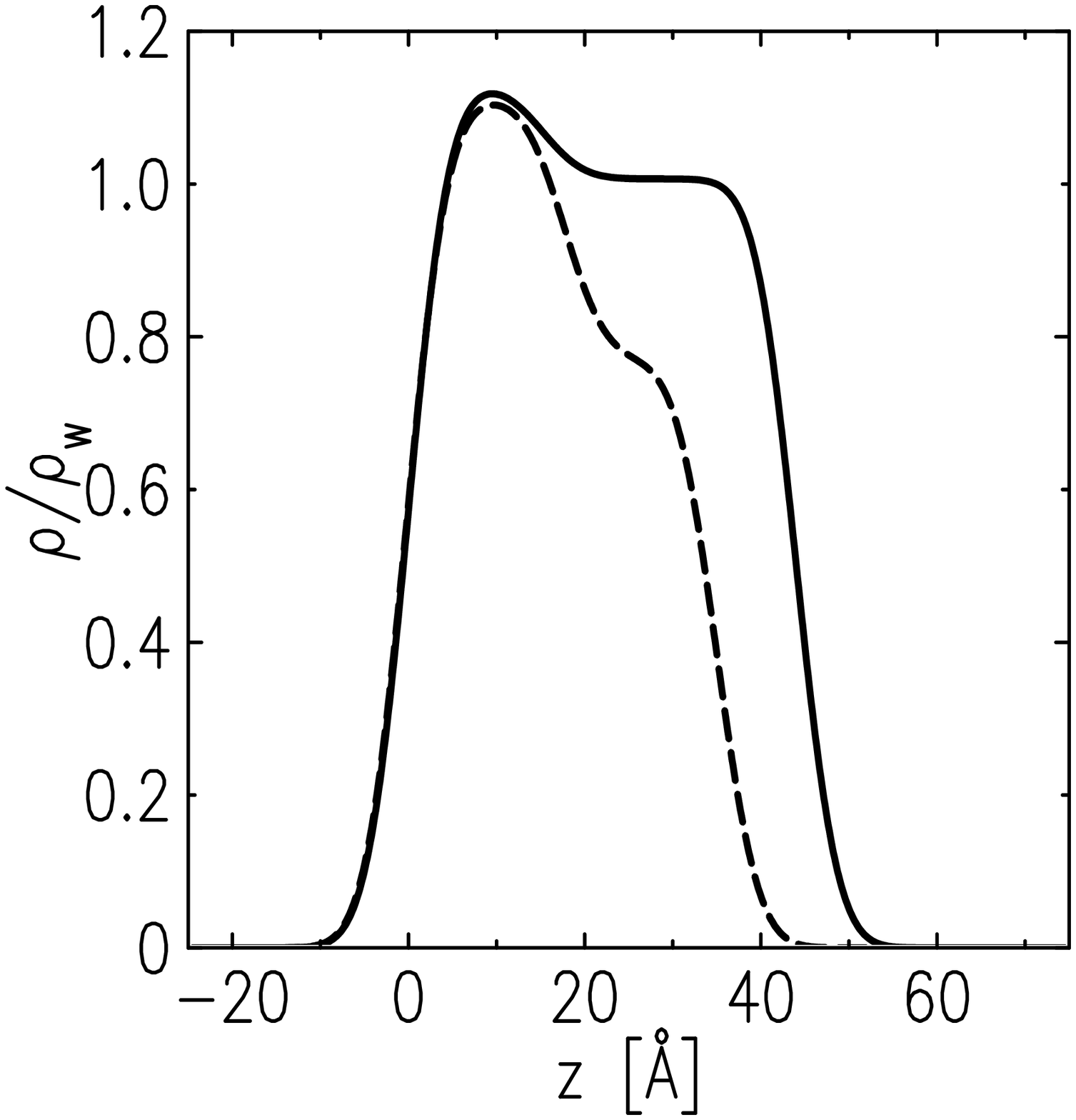, width=0.7\textwidth}

Figure 6. Model electron density profiles for a monolayer of
melissic acid normalized to the electron density in water
under normal conditions ($\rho_w=0.333$\,{\it e$^-$/}{\AA}$^3$) obtained
within the two-layer model of (solid line) the low-temperature
phase of the monolayer ($T = 293.4$\,K) and
(dashed line) the high-temperature phase of the monolayer
($T = 317.9$\,K).
\end{figure}

The analysis of Figs. 3 and 4 shows that C$_{30}$H$_{60}$O$_2$
molecules are absorbed in the form of a monolayer on
the n-hexane -– water interface. In order to obtain
detailed information on the structure of the interface
from $R(q_z)$, we used the simplest qualitative two-layer
model (slab model) of an adsorbed layer with five fitting
parameters in which the density profile $\rho(z)$ is
based on the error function \cite{14}. The lower limit of the
standard deviations $\sigma_j$ of the positions of the $j$th interfaces
of the bilayer ( j = 0, 1, 2) from the reference
value $z_j$ is determined by the capillary width $\sigma_{cw}^2 = ( k_BT/2\pi\gamma )\ln(Q_{max}/Q_{min})$ ($k_B$ is the Boltzmann constant),
which is specified by the short-wavelength limit
in the spectrum of capillary waves $Q_{max} = 2\pi/a$ (where
$a\approx 10$\,{\AA} is the order of magnitude of the molecular
radius) and $Q_{min}=q_z^{max}\Delta\beta$ (where $q_z^{max} \approx 0.45$ 
{\AA}$^{-1}$)\cite {15,16,17}. Under the assumption that $\sigma_j=\sigma_0$ for all $j$ 
values, the structure factor of the surface in the first Born
approximation has the form \cite{18}
\begin{equation}
\frac{R(q_z)}{R_F(q_z)} \approx  \left| \frac{1}{\Delta\rho}\sum_{j=0}^{2}{(\rho_{j+1}-\rho_j) e^{-iq_zz_j}} \right|^2 e^{-\sigma_0^2q_z^2},
\end{equation}
where $\rho_j$ is the electron density in the $j$th layer; $\rho_0$ and $\rho_3$
are the electron densities in water and n-hexane, respectively; and $R_F(q_z)\approx$ $(q_z-[q_z^2-q_c^2]^{1/2})^2/(q_z+[q_z^2-q_c^2]^{1/2})^2$
is the Fresnel function. The calculated reflection curves are shown
by solid lines in Figs. 3 and 4. The model profiles $\rho(z)$
for monolayers of (solid line) low- and (dashed line)
high-temperature phases are shown in Fig. 6.

The variation of the parameters in the model of the
monolayer is in agreement with the molecular structure
of melissic acid, which has a hydrophilic head
part and a hydrophobic hydrocarbon tail. The first
hydrophilic layer of the monolayer for the low-temperature
phase, which is in direct contact with water,
includes –ÑÎÎÍ polar groups and has the density $\rho_1=(1.16\pm0.05)\rho_w$
and the thickness $L_1=z_1-z_0=(15\pm2)$\,\AA{} 
 ($\rho_w=0.333$\,{\it e$^-$/}{\AA}$^3$ is the electron density in
water under normal conditions). If the width of this
layer is fixed $L_1 < 10$\,\AA{} at fitting, the quality of fitting
at high $q_z$ values worsens significantly. The second
layer with the thickness $L_2 =z_2-z_1= (29\pm2)$\,\AA{} is
formed by hydrocarbon chains with the density $\rho_2=(1.01\pm0.02)\rho_w$.
 The fitting parameter $\sigma_0$ varies from 3.6\,\AA{} to 4.5\,\AA{}, 
which coincides within the errors with the
calculated $\sigma_{cw}$ value. The total thickness of the monolayer
is $(42\pm 3)$\,\AA{}, which also coincides within the
errors with the calculated total length (40.8\,\AA{} 
$=29\times 1.27$\,\AA (C-C) + 1.5\,\AA (-CH$_3$) + 2.5\,\AA (-COOH)) of the
C$_{30}$H$_{60}$O$_2$ molecule (it contains 256 electrons). Thus, all 
molecules in this phase of the monolayer are elongated
along the normal to the surface and the area per
molecule of C$_{30}$-acid is $A= 256/(\rho_1L_1+\rho_2 L_2)$ $= (17\pm 1)$\,\AA$^2${}. 
This value corresponds to the densest crystal
phase of alkanes \cite{19}.

The hydrophilic layer in the high-temperature
phase has the density $\rho_1=(1.12\pm0.02)\rho_w$ and thickness
$L_1=(18\pm2)$\,\AA{}. The electron density in the second
layer with the thickness $L_2=(18\pm2)$\,\AA{} is $\rho_1=(0.77\pm 0.02)\rho_w$.
 The fitting value $\sigma_0 = 4.0\pm 0.2$\,\AA{}
coincides within the errors with the value $\sigma_{cw}\approx 3.6$\,\AA{}
calculated with the data for $\gamma(T)$. The value $A=23\pm 1$\,\AA$^2$ 
for the high-temperature phase corresponds to a
high-molecular weight hydrocarbon liquid \cite{19}.

The monotonic temperature dependence $R(T)$ at a
fixed $q_z$ value (see Fig. 7) indicates incoherent reflection
from the in-plane structure of the interface.
Since our data show that the relative contribution from
diffuse scattering at the interface to the reflected
power is small $\sim 10^{-3}$, $R$ in the first approximation
can be represented in the form of a linear function of
the coverage of the surface $C(T)$ by domains of the low-temperature
phase of C$_{30}$-acid \cite{20}:
\begin{equation}
R\approx C(T)R_1+(1-C(T))R_2,
\end{equation}
where $R_1$ and $R_2$ are the $R$ values for low-temperature
($C(T) = 1$) and high-temperature ($C(T) = 0$) phases,
respectively. Circles in Fig. 7 represent the calculated
dependence for $C(T)\approx (R-R_2)/(R_1-R_2)$ at $q_z=0.05$\,\AA$^{-1}$.
 According to this dependence, the phase
transition in the monolayer at $T_c\approx 293.5$\,Ê is sharp:
the surface is rearranged in the temperature interval
$\leq 0.5$\,Ê. A small difference ($\sim 2$\,K) of the $T_c$ value for
this sample from the temperature of the bending point
$\gamma(T)$ in Fig. 2 is due both to uncertainty in its determination
and to a small difference in the volume concentration
$c$ between the samples.

The analysis of the experimental results reveals two
important features of the critical behavior of the crystalline
monolayer of melissic acid at the interface.
First, a phase transition associated with melting
occurs in this monolayer with an increase in the temperature.
Such a behavior differs from the critical
behavior of monolayers of high-molecular saturated
and fluorocarbon alcohols at this interface, where an
increase in the temperature is accompanied by phase
transitions of evaporation and sublimation of monolayers,
respectively \cite{6}. Second, according our data,
the total electron density in the monolayer at $T_c$
decreases stepwise by $\Delta A/A\approx 30\%$. This behavior is
significantly different from the critical behavior of
monolayers of high-molecular weight alcohols at the n-hexane –-
water interface, where phase transitions occur in
a certain temperature range, as well as from the crystallization
of monolayers of CTAB and STAB cationic
surfactants, where two critical temperatures were
observed \cite{7}.

\begin{figure}
\hspace{0.5in}
\epsfig{file=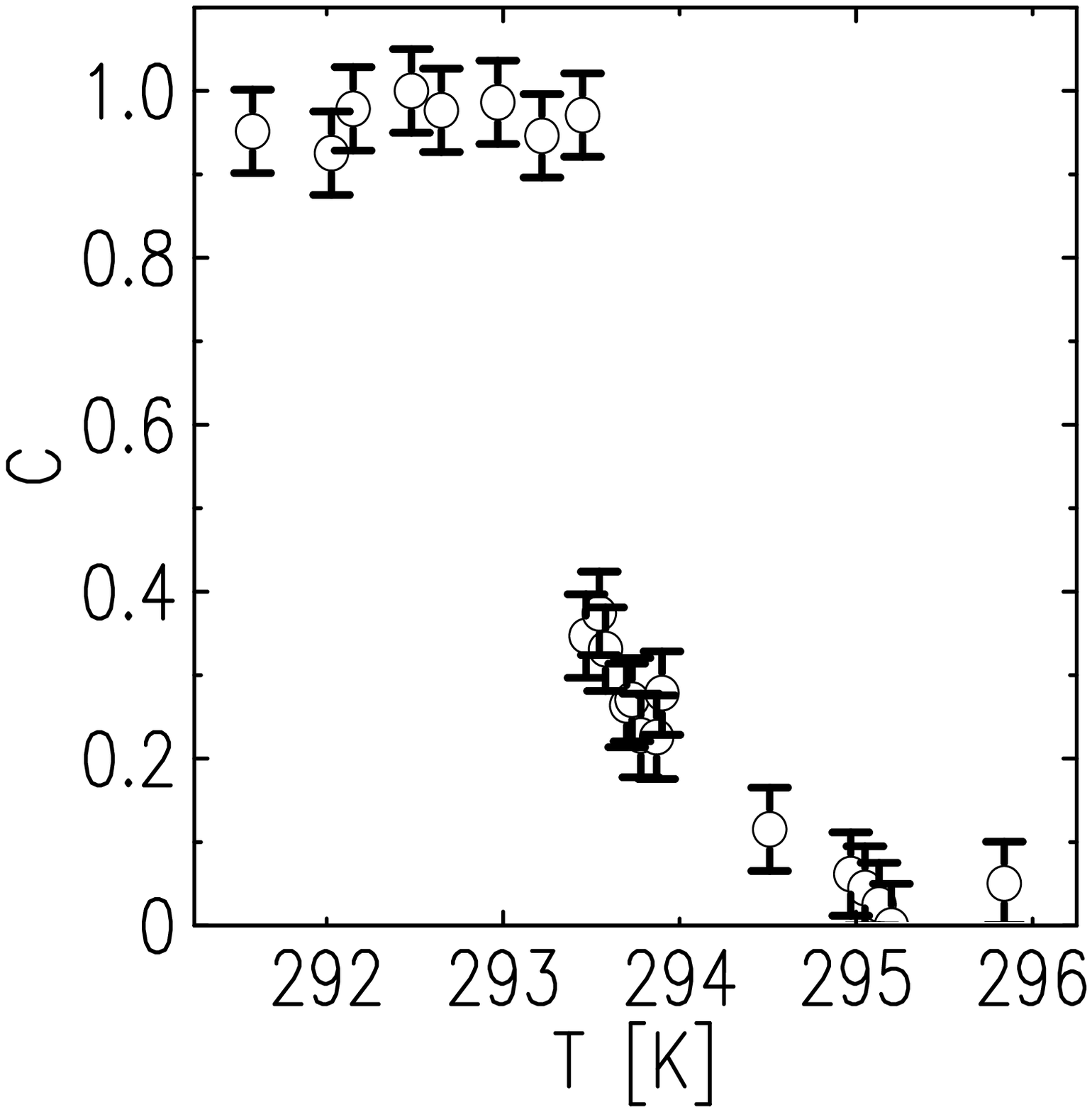, width=0.7\textwidth}

Figure 7. Temperature dependence of the coverage of the interface
by domains of the low-temperature phase $C$. The
points are obtained from the data presented in Fig. 5 with
the use of Eq. (2), where $R_1$ and  $R_2$ are the reflection coefficients
at $T\approx 292$\,K and $T\approx295$\,K, respectively.
\end{figure}

To conclude, the reported data illustrate the solid -–
liquid phase transition in the Gibbs monolayer at the
n-hexane -– water interface. With an increase in the
temperature in a narrow vicinity of $T_c$, a significant
fraction of adsorbed molecules of C$_{30}$-acid leave the
interface and are solved in the bulk of n-hexane. In
this case, the thickness of the monolayer $L_1 + L_2$
decreases by $\approx 15\%$ and $A$ increases by $\approx 30\%$.

I am grateful to Prof. Mark L. Shlossman and Prof. Vladimir I. Marchenko
for stimulating discussions of the experimental
results. The work at the NSLS synchrotron was supported
by the US Department of Energy (contract no.
DE-AC02-98CH10886). The work at the X19C station
was supported by the ChemMatCARS Foundation,
University of Chicago, University of Illinois at
Chicago, and Stony Brook University.

\end{document}